\documentclass[12pt]{article}
\usepackage{amsfonts}
\usepackage{amsmath}
\usepackage{amssymb}
\usepackage{graphicx}
\usepackage{color}
\usepackage[all, knot]{xy}
\usepackage{tikz}
\usepackage{array}
\usepackage{hyperref}
\usepackage{ulem}

\usepackage{comment}

\usepackage[utf8]{inputenc}
\usepackage{epstopdf}
\usepackage[footnotesize]{caption}
\usepackage{subcaption}
\usepackage{amsthm}
\usepackage{enumitem}
\usepackage{mathrsfs}
\usepackage{mathtools}

\usepackage[margin=3cm]{geometry}

\def \be {\begin{equation}}
\def \ee {\end{equation}}
\def \bea {\begin{eqnarray}}
\def \eea {\end{eqnarray}}
\def \nn {\nonumber}

\def \rr {\raise.35ex\hbox{\small $\prime$}\kern-.17em{\mbox{\large $\imath$}}}

\def \dels {\partial\kern-.6em /\kern.1em}
\def \As {{A\kern-.5em / \kern.5em}}
\def \Ds {D\kern-.7em / \kern.5em}

\def \ks {k\kern-.5em /}
\def \ls {l\kern-.5em /}

\newcommand{\ci}[1]{}

\newcommand{\ba}{\begin{eqnarray}}
\newcommand{\ea}{\end{eqnarray}}
\newcommand{\bal}{\begin{align}}
\newcommand{\eal}{\end{align}}
\newcommand{\bay}[1]{\left(\begin{array}{#1}}
\newcommand{\eay}{\end{array}\right)}

\newcommand{\ket}[1]{|{#1}\rangle}

\def\mf{{m_\mathrm{f}}}

\newcommand{\hide}[1]{}

\newlist{axioms}{enumerate}{2}
\setlist[axioms,1]{label=\textbf{A\arabic{axiomsi}.}, ref=A\arabic{axiomsi}}
\setlist[axioms,2]{label=\textbf{A\arabic{axiomsi}\rlap{\myEnumCounter{axiomsii}}.},%
                   ref=A\arabic{axiomsi}\myEnumCounter{axiomsii},%
                   align=parleft,%
                   leftmargin=0em,%
                   itemsep=1.4ex,%
                   before={\stepcounter{axiomsi}}}

  \usetikzlibrary{decorations.markings}

\begin{document}

\begin{titlepage}
\begin{center}

\textbf{\LARGE
Matter Coupling of Dirac Matter in the Context of the SYK Model: Non-Gaussian Random Couplings and Bulk Mass Deformations
\vskip.3cm
}
\vskip .5in
{\large
Pak Hang Chris Lau$^{a,b}$ \footnote{e-mail address: phcl2@gbu.edu.cn},
Chen-Te Ma$^{a}$ \footnote{e-mail address: yefgst@gmail.com}, 
Jeff Murugan$^{c,d}$ \footnote{e-mail address: jeff.murugan@uct.ac.za}, 
and\\ 
Masaki Tezuka$^e$ \footnote{e-mail address: tezuka@scphys.kyoto-u.ac.jp} 
\\
\vskip 1mm
}
{\sl
$^a$
Department of Physics, Great Bay University, Dongguan, Guangdong 52300, China. 
\\
$^b$
Department of Physics, Osaka University, Toyonaka, Osaka 560-0043, Japan.
\\
$^c$
The Laboratory for Quantum Gravity and Strings,\\
Department of Mathematics and Applied Mathematics,
University of Cape Town, Private Bag, Rondebosch 7700, South Africa.
\\
$^d$
National Institute for Theoretical and Computational Sciences, 
Private Bag X1, Matieland,
South Africa.
\\
$^e$ 
Department of Physics, Kyoto University, Kitashirakawa, Sakyo-ku, Kyoto 606-8502, Japan.
}
\\
\vskip 1mm
\vspace{40pt}
\end{center}
\newpage
\begin{abstract}
\noindent
We elaborate further on the matter coupling of Dirac matter in the SYK framework, incorporating non-Gaussian coupling distributions and bulk fermion mass effects.
Our study analyzes quartic matter couplings generated by a non-Gaussian distribution as an illustrative example.
The introduction of bulk-fermion mass alters the boundary coupling between the Dirac and Majorana fermions.
The averaged adjacent gap ratio is sensitive to the distribution of random couplings, which remains independent of the Hamiltonian's symmetry.
The generalization of the SYK model to non-Gaussian distributions and the inclusion of bulk fermion mass remain qualitatively similar to the Gaussian and massless cases.
Key deviations are observed only in the time scales for the linear ramp in the spectral form factor and the saturation of entanglement entropy.
\end{abstract}
\end{titlepage}

\section{Introduction}
\label{sec:1}
\noindent
Jackiw–Teitelboim (JT) gravity -- a two-dimensional gravitational theory coupled to a dilaton -- offers a compelling framework for exploring quantum aspects of gravity, difficult or impossible to access in higher dimensions \cite{Teitelboim:1983ux,Jackiw:1984je}.
In 2D, the Einstein tensor
\bea
G_{\mu\nu}\equiv R_{\mu\nu}-\frac{1}{2}g_{\mu\nu}R,
\eea
vanishes identically, making the dynamics entirely dependent on the dilaton field and the cosmological constant. With a negative cosmological constant, the boundary dynamics of JT gravity \cite{Jensen:2016pah,Maldacena:2016upp,Engelsoy:2016xyb,Bagrets:2016cdf} have connections to conformal field theory and the Sachdev--Ye--Kitaev (SYK) model \cite{Polchinski:2016xgd}.
JT gravity plays a crucial role in the two-dimensional manifestation of the Anti-de Sitter/Conformal Field Theory (AdS/CFT) correspondence in which the bulk geometry is a patch of AdS$_2$, characterized by constant negative curvature $R= 2\Lambda$ with $\Lambda <0$ and metric
\bea
ds^2=-\frac{1}{\Lambda}\frac{dt^2+dz^2}{z^2}\,,
\eea
with $z\rightarrow 0$ defining the conformal boundary.
This AdS$_2$ theory is dual to a one-dimensional quantum mechanics related to the SYK model, a solvable quantum mechanical model of fermions with random interactions.
Among its very many remarkable properties, which have been well-documented in both the high-energy and condensed matter literature, the SYK model exhibits an {\it emergent} conformal symmetry at low energies \cite{Polchinski:2016xgd}, a necessary condition for any presumptive dual to JT gravity \cite{Maldacena:2016hyu}.
The low-energy boundary degrees of freedom in JT gravity reduce to the {\it Schwarzian} action \cite{Jensen:2016pah,Maldacena:2016upp,Engelsoy:2016xyb,Bagrets:2016cdf},
\bea
S_{\mathrm{SCH}}=-\frac{1}{8\pi G_2}\int^{\infty}_{-\infty}du\
\phi_b\bigg(\frac{f^{\prime\prime\prime}}{f^{\prime}}-\frac{3}{2}\frac{f^{\prime\prime 2}}{f^{\prime 2}}\bigg)\,,
\eea
where $\phi_b$ is the value of the dilaton on the boundary, and $G_2$ is the Newton constant in two dimensions.
Promoting the reparametrization function $f$ to a boundary field is successful because the solution space of bulk and boundary theories is equivalent \cite{Maldacena:2016upp}.
\\

\noindent
On the 1-dimensional boundary, the Hamiltonian for the $\mathrm{SYK}$ model describes the quantum mechanics of $N$ Majorana fermions with random four-body interactions.
The SYK model offers an insightful framework for bridging concepts from holography and quantum chaos.
Its chaotic nature is demonstrated through the statistical behavior of its energy spectrum, which is aligned with predictions from Random Matrix Theory (RMT) \cite{Dyson:1962es}.
The model's energy level distribution, analyzed using metrics such as the adjacent gap ratio \cite{Atas:2012prl,Nishigaki:2024yjr} and the spectral form factor \cite{Brezin:1997rze,Lau:2018kpa,Lau:2020qnl}, confirms its alignment with RMT predictions \cite{Garcia-Garcia:2016mno,Dyer:2016pou,Cotler:2016fpe,Krishnan:2016bvg}.
This behavior is characteristic of many-body quantum chaotic systems, which universally exhibit RMT spectra in the semi-classical limit \cite{Muller:2004nb}.
The SYK model provides a concrete model where holographic duality and quantum chaos coexist, highlighting how chaotic phenomena serve as diagnostic tools for holographic systems.
The universality of RMT in chaotic systems is pivotal in establishing this link.
\\

\noindent
While it has become increasingly clear that quantum chaos plays a central role in the emergence of spacetime, the precise mechanism by which gravitational dynamics arise from underlying quantum systems remains elusive. Recent advances suggest that von Neumann (vN) algebras may provide a crucial structural framework for uncovering this mechanism \cite{Leutheusser:2021qhd}.
As classifications of operator algebras acting on Hilbert spaces, vN algebras offer a natural language for distinguishing between quantum systems with different informational and dynamical properties — including those that exhibit chaotic behavior.
In particular, gravitational features emerge from the transition from Type I algebras, typical of finite-dimensional quantum systems, to Types II and III, which are more appropriate for quantum field theories and holographic gravity.
This interplay hints at a deeper connection between the algebraic structure of operator growth in chaotic systems and the emergence of spacetime geometry.
\\

\noindent
The distinction between von Neumann algebra types provides a powerful lens through which to view the transition from quantum gravity to semi-classical regimes, ultimately to pure gravity.
In this context, the SYK model offers a fertile testing ground for exploring the emergence of spacetime.
At finite $N$, the SYK model is described by a Type I von Neumann algebra, consistent with its realization as a finite-dimensional quantum system.
However, in the large-$N$ limit, its operator structure begins to resemble that of Type II or III algebras, which are more appropriate for quantum field theories and gravitational systems \cite{Lau:2023pot}.
The defining feature of Type III algebras — the absence of well-defined density matrices — is not a bug but a feature: it captures the physics of systems where entropy and temperature are emergent rather than fundamental, as in the case of event horizons.
This algebraic structure aligns naturally with holographic dualities, where spacetime geometry emerges from entanglement patterns in boundary quantum systems.
\\

\noindent
The SYK model, with its strongly interacting, disordered structure, again serves as a particularly clean and tractable laboratory for investigating these phenomena.
As its large-$N$ limit captures features reminiscent of semi-classical gravity, it offers a concrete realization of how collective quantum behavior might give rise to gravitational dynamics.
Through the lens of holography — particularly the AdS/CFT correspondence — such models provide valuable insights into how matter and spacetime might emerge from a quantum gravitational substrate \cite{Lau:2023pot}.
\\

\noindent
The solvability of the SYK model stands in contrast to the challenges associated with understanding its bulk gravitational dual, particularly due to the non-local features inherent in the bulk theory \cite{Gross:2017hcz}.
This underscores a central difficulty in using the 1/$N$ expansion to bridge the SYK boundary theory with a bulk semi-classical description.
While the 1/$N$ expansion captures the complete semi-classical limit to all orders and aligns with expectations from large-$N$ holographic dualities, it notably lacks non-perturbative corrections.
As a result, this framework may not include essential quantum gravitational phenomena—including those relevant to the black hole information paradox.
\\

\noindent
This observation suggests that even semi-classical gravity must encode some resolution of the information paradox at energy scales below certain thresholds.
A key insight is that any effective bulk theory with emergent gravitons must necessarily involve non-trivial couplings to matter.
Understanding the structure and consequences of these couplings is essential for uncovering semi-classical mechanisms that could mitigate or resolve information loss.
The interplay between graviton dynamics and matter couplings is particularly fertile ground for further investigation, especially within models inspired by SYK \cite{Almheiri:2014cka}.
\\

\noindent
Modeling bulk matter fields from the boundary CFT remains a significant challenge.
One promising recent approach introduces random Yukawa-SYK couplings to incorporate scalar fields into the Hamiltonian \cite{Breitenlohner:1982bm,Anninos:2020cwo,Moitra:2022glw,Anninos:2022qgy}.
The tunable parameter controlling the distribution of random couplings offers a systematic approach to handling coupling effects, enhancing our understanding of scalar dynamics in large-$N$ settings.
Extending this framework to include fermionic matter and employing numerical methods to study the resulting dynamics has proven promising \cite{Lau:2023pot}.
Because the 1D fermionic matter coupling corresponds to the degrees of 2D bulk Dirac fermions, we call this fermion matter a Dirac fermion.
Since the modification term does not correspond to the coupling between the dilaton and matter fields in the large $N$ limit, our matter coupling does not affect the JT background and the SYK's conformal solution under the perturbation context but introduces the detailed modification of the large-$N$ model. 
Furthermore, introducing non-Gaussian random couplings \cite{Peng:2022pfa} allows for the generation of quartic interactions, enriching the model and opening new avenues for exploring emergent matter-gravity interactions.
\\

\noindent
Building on our previous work \cite{Lau:2023pot}, this paper provides a comprehensive exploration of JT gravity coupled to Dirac matter through the SYK model, introducing several interesting generalizations—such as the introduction of fermion mass and consideration of non-Gaussian disordered averages—and analyzing their effects.
In this sense, our work bridges the SYK model's deformation and JT gravity, focusing on fundamental symmetry properties and dynamical features such as time scales. Specifically;
\begin{itemize}
\item
We derive the boundary description of JT gravity with Dirac matter coupling, adjusting the SYK model's deformation to remain consistent with the corresponding gravitational theory.

\item
Despite introducing quartic matter couplings, the symmetry of the Hamiltonian remains intact. While the adjacent gap ratio reveals that a non-Gaussian disorder distribution impacts spectral statistics, the fundamental symmetry classification is preserved at sufficiently large $N$. However, since non-Gaussianity is not suppressed in the large-$N$ limit, the standard symmetry classifications inherited from the Gaussian case become non-trivial and deserve careful reinterpretation.

\item
The adjacent gap ratio probes short-range spectral correlations, confirming the stability of the symmetry classification.
In contrast, the spectral form factor (SFF) captures long-range correlations. Notably, non-Gaussianity extends the linear ramp's timescale or, with bulk fermion mass, reduces it in the SFF, providing a potentially measurable signature of this generalization.

\item
An entanglement entropy analysis reveals maximal entanglement between graviton and Dirac matter in equilibrium, signaling a transition from a Type I to a Type II$_{1}$ von Neumann algebra as $N \to \infty$. Non-Gaussian effects and finite fermion mass modify the entanglement saturation timescale, although the large-$N$ equilibrium entanglement structure remains robust. However, including all non-Gaussian contributions and general bulk-matter couplings can destabilize this equilibrium. In particular, additional bulk-matter couplings can disrupt the maximally mixed state, suggesting the emergence of a Type II$_{\infty}$ algebra on intermediate timescales. We, therefore, indicate that truncations enforcing a Type II$_1$ structure or rapid entanglement saturation should be viewed as modeling artifacts rather than fundamental features.

\item
Neither non-Gaussianity nor finite bulk fermion mass modifies the random matrix symmetry classification or the late-time maximally entangled equilibrium in the large-$N$ limit. Their observable effects are confined to the timescales associated with the SFF linear ramp and the approach to saturation of entanglement entropy.
\end{itemize}

\noindent
The organization of the rest of this paper is as follows: In Sec.~\ref{sec:2}, we provide a detailed analysis of the boundary description of JT gravity, incorporating the matter coupling from the low-energy theory of the SYK model and the effects of Dirac fermions.
Sec.~\ref{sec:3} focuses on the computation of the adjacent gap ratio, the spectral form factor, and the entanglement entropy.
Finally, Sec.~\ref{sec:4} concludes with a summary of our findings and outlook for future work.


\section{JT Gravity, the SYK Model and Matter Coupling}
\label{sec:2}
\noindent
We begin by reviewing JT gravity \cite{Teitelboim:1983ux,Jackiw:1984je} and its coupling to matter through quadratic matter-coupling terms.
We then derive the corresponding boundary descriptions.
In particular, by introducing appropriate interaction terms between Majorana fermions and matter fields in the SYK model, effective low-energy theories can be obtained that are consistent with bulk gravitational dynamics.
Furthermore, quartic matter-coupling terms can be incorporated through a non-Gaussian disorder average without explicitly modifying the microscopic Hamiltonian.

\subsection{JT Gravity and Matter Coupling}
\noindent
We first introduce the setting of pure JT gravity.
The treatment of matter coupling, whether from scalar or fermionic fields, follows a parallel structure.
We will present the boundary description for the scalar case explicitly, then summarize the extension to Dirac matter. \noindent
The JT gravity action, $S_{\mathrm{JT}}$, consists of a bulk and a boundary contribution,
\begin{eqnarray}
S_{\mathrm{JT}} &=& -\frac{1}{16\pi G_2}\int_{\mathcal{M}} d^2x\, \sqrt{|\det g_{\mu\nu}|}\, \phi (R - 2\Lambda)\nonumber\\
&& - \frac{1}{8\pi G_2}\int_{\partial \mathcal{M}} du\, \sqrt{|\det h_{uu}|}\,\phi K\,
\end{eqnarray}
where the dilaton field $\phi$ couples linearly to the Ricci scalar $R$, enforcing constant curvature via its equations of motion. The boundary term ensures a well-posed variational principle under Dirichlet boundary conditions, which fixes the boundary data on $\partial \mathcal{M}$.
Here, $h_{uu}$ denotes the induced boundary metric, while the extrinsic curvature is given by,
\bea
K \equiv g^{\mu\nu}\nabla_{\nu} n_{\mu}\,,
\eea
with $n^{\mu}$ a unit normal vector satisfying,
\bea
n^{\rho} n_{\rho} = 1\,,
\eea
assuming a negative cosmological constant, $\Lambda < 0$.

\subsubsection{Scalar field coupling}
\noindent
For scalar matter, the simplest coupling terms consist of a kinetic term and a mass term:
\bea
S_{\Phi}
&=&\frac{\lambda}{32\pi G_2\widetilde{M}}\sum_{k=1}^{\widetilde{M}}\int d^2x\sqrt{|\det{g_{\rho\sigma}}|}\ (\nabla_{\mu}\Phi_k)(\nabla^{\mu}\Phi_k)
\nn\\
&&
+\frac{\lambda m^2}{32\pi G_2\widetilde{M}}\sum_{k=1}^{\widetilde{M}}\int d^2x\ \sqrt{|\det g_{\rho\sigma}|}\ \Phi_k\Phi_k\,,
\eea
where $\lambda$ is the matter coupling strength, $\widetilde{M}$ counts the number of scalar fields, and $m$ is their mass.
In AdS spacetime, the conformal dimension of these scalars is,
\bea
\Delta=\frac{1}{2}+\sqrt{\frac{1}{4}+\frac{m^2}{\Lambda}}\,.
\eea
The resulting equation of motion is given by,
\bea
\bigg(\partial_+\partial_- -\frac{1}{2}e^{2\rho}m^2\bigg)\Phi_k=0\,,
\eea
where
\bea
e^{2\rho}=\frac{1}{2\Lambda z^2}\,, \quad x^{\pm}=t\pm iz\,.
\eea
Near the boundary, the field behaves as
\bea
\lim_{z \to 0} \Phi_k(x^+, x^-) = z^{1-\Delta} j_k(t)\,,
\eea
with source $j_k$.
The general solution is
\bea
\Phi_k(x^+, x^-) = c \int_{-\infty}^{\infty} d\tau\ \frac{z^{\Delta}}{\left[(\tau - t)^2 + z^2\right]^{\Delta}} j_k(\tau)\,,
\eea
where
\bea
\frac{1}{c} = \int_{-\infty}^{\infty} dy\ \frac{1}{(1+y^2)^{\Delta}}\,.
\eea
The resulting boundary action is
\bea
S_{\mathrm{bds}} &=& -\frac{\lambda}{32\pi G_2 \widetilde{M}} \sum_{k=1}^{\widetilde{M}} \int dt\ \Phi_k \partial_z \Phi_k
\nn\\
&=& -\frac{c \lambda}{64\pi G_2 \widetilde{M}} \int_{-\infty}^{\infty} d\tau_1 d\tau_2\
\frac{j_k(\tau_1) j_k(\tau_2)}{|\tau_1 - \tau_2|^{2\Delta}}.
\eea

\subsubsection{Dirac matter coupling}
\noindent
Next, we consider the coupling of the Dirac matter with a mass $m_f$. This case is structurally similar to the scalar field. Following the treatment above:
\bea
S_{\Psi}&=&\frac{\lambda}{32\pi G_2\widetilde{M}}\sum_{k=1}^{\widetilde{M}}\int d^2x\sqrt{|\det g_{\rho\sigma}|}\
\bar{\Psi}_k\bar{\gamma}^{\mu}\overleftrightarrow{D}_{\mu}\Psi_k
\nn\\
&&
+\frac{\lambda \mf}{16\pi G_2\widetilde{M}}\sum_{k=1}^{\widetilde{M}}\int d^2x\sqrt{|\det g_{\rho\sigma|}}\ \bar{\Psi}_k\Psi_k,
\eea
where
\bea
\bar{\Psi}_k\equiv\Psi_k^{\dagger}\gamma^0 \,,\qquad \bar{\gamma}^{\mu}\equiv e_a{}^{\mu}\gamma^a.
\eea
The vielbein indices are denoted by Latin letters $a,b,\cdots$, and $\mf$ is the mass of the bulk fermion fields, $\Psi_k$. Dirac $\gamma$-matrices in the vielbein basis are given by,
\bea
\gamma_0\equiv\sigma_x=\begin{pmatrix}
0&1
\\
1&0
\end{pmatrix}, \
\gamma_1\equiv-\sigma_y=\begin{pmatrix}
0&i
\\
-i&0
\end{pmatrix}\,,
\eea
and the derivative operator $\overleftrightarrow{D}_{\nu}$ acts as
\bea
\bar{\Psi}_k\bar{\gamma}_{\mu}\overleftrightarrow{D}_{\nu}\Psi_k=
\bar{\Psi}_k\bar{\gamma}_{\mu}\overrightarrow{D}_{\nu}\Psi_k-\bar{\Psi}_k\overleftarrow{D}_{\nu}\bar{\gamma}_{\mu}\Psi_k,
\eea
where
\bea\overrightarrow{D}_{\mu}\equiv\overrightarrow{\partial}_{\mu}+\frac{1}{8}\eta_{ca}\omega_{\mu}{}^c{}_b\lbrack\gamma^a, \gamma^b\rbrack; \qquad
\overleftarrow{D}_{\mu}\equiv\overleftarrow{\partial}_{\mu}-\frac{1}{8}\eta_{ca}\omega_{\mu}{}^c{}_b\lbrack\gamma^a, \gamma^b\rbrack,
\eea
with spin connection
\bea
\omega_{\mu}{}^a{}_b=-e_b{}^{\nu}(\partial_{\mu}e^a{}_{\nu}-\Gamma^{\lambda}_{\mu\nu}e^a{}_{\lambda})\,.
\eea
The conformal dimension of the fermion is
\bea
\Delta=\frac{1}{2}+\frac{\mf}{\sqrt{-\Lambda}}\,.
\eea
The boundary description for the coupling of Dirac matter is given by
\bea
-\frac{\lambda}{16\pi G_2\widetilde{M}}\sum_{k=1}^{\widetilde{M}}\int dt\ \bar{\Psi}_k\bar{\gamma}_z\Psi_k
\sim
-\frac{\lambda}{16\pi G_2\widetilde{M}}\sum_{k=1}^{\widetilde{M}}\int d\tau_1d\tau_2\ \frac{J_k^{\dagger}(\tau_1)J_k(\tau_2)}{|\tau_1-\tau_2|^{2\Delta}},
\eea
in which $\sim$ denotes agreement up to an overall constant.
The boundary field is
\bea
J_k\equiv\begin{pmatrix}
j_{1, k}
\\
j_{2, k}
\end{pmatrix}.
\eea
\noindent
In general, the boundary description for both scalar and fermionic matter couplings can be compactly expressed as
\bea
S_{\mathrm{bdt}}
&=& -\frac{1}{8\pi G_{2}} \int_{-\infty}^{\infty} \!\!du\,\,
\phi_b\bigg(\frac{f^{\prime\prime\prime}}{f^{\prime}}-\frac{3}{2}\frac{f^{\prime\prime 2}}{f^{\prime 2}}\bigg)
\nn\\
&& - b \frac{\lambda}{16\pi G_2 \widetilde{M}} \sum_{k=1}^{\widetilde{M}} \int_{-\infty}^{\infty} du_1 du_2\
\frac{f’(u_1) f’(u_2)}{ |f(u_1) - f(u_2)|^{2\Delta} } \tilde{J}_k^{\dagger}(u_1) \tilde{J}_k(u_2),
\eea
where $b$ is a constant and $\tilde{J}_k$ is the reparametrized boundary source field. Promoting $f$ from a function to a dynamical field is justified because its variation consistently captures the solution space of the bulk gravitational theory.
The result reflects that it only serves as a re-expression within the JT gravity framework, illustrating how reparametrization invariance acts on the correlation function.

\subsection{SYK Model and Matter Coupling}
\noindent
Having established the boundary description of JT gravity with scalar and fermionic matter coupling, we now turn to the microscopic SYK realization.
We introduce matter coupling into the SYK model in a manner consistent with the boundary description of JT gravity.
The low-energy effective theory of the SYK model \cite{Polchinski:2016xgd} is governed by the Schwarzian action, which captures the same boundary dynamics as pure JT gravity \cite{Jensen:2016pah,Maldacena:2016upp,Engelsoy:2016xyb,Bagrets:2016cdf}.
To reproduce the quadratic coupling terms derived in the JT framework, we must incorporate suitable interaction terms between the Majorana fermions and additional matter fields within the SYK model.
We will argue that such quartic matter couplings can emerge dynamically via a non-Gaussian disorder average, thereby enriching the effective theory without requiring explicit microscopic couplings.
Our perturbation analysis in the Gaussian disorder average is only applicable in the weak non-Gaussianity regime of which the result should align with the Gaussian case as the coupling constant approaches zero. 
We will analysis the strong coupling regime using numerical analysis in next section where significant deviation from perturbative analysis is observed. 
This formulation provides a consistent framework for analyzing how gravitational features — including the interplay of dilaton dynamics, matter coupling, and emergent reparametrization modes — can be systematically studied within a tractable many-body quantum model.
As it was pointed out in Ref. \cite{Moitra:2022glw}, a 1D bosonic field with a canonical kinetic term violates the unitarity bound. 
We follow their strategy in this paper and limit our discussion only to the long time or low energies regime where $gt\gg1$ and $g/J^{3/2}\ll1$. 
In this particular regime, the kinetic term of the scalar matter can be safely neglected while keeping our perturbative analysis valid. 
In our numerical analysis or the next section, our focus is on the fermionic matter where this subtlety does not arise. 
In the following sections, we explicitly construct the relevant deformations of the SYK Hamiltonian, compute their low-energy consequences, and analyze their holographic interpretation.

\subsubsection{SYK Model}
\noindent
In its most vanilla form, the SYK model is defined through the Hamiltonian \cite{Polchinski:2016xgd},
\begin{eqnarray}
H_{\mathrm{SYK}} = \sum_{1 \le i_1 < i_2 < i_3 < i_4 \le N} J_{i_1 i_2 i_3 i_4} \psi_{i_1} \psi_{i_2} \psi_{i_3} \psi_{i_4}\,,
\end{eqnarray}
where the $\psi_i$ are Majorana fermion operators that satisfy the anticommutation relations $\{\psi_i, \psi_j\} = \delta_{ij}$, and $J_{i_1 i_2i_3 i_4}$ are random coupling constants that are drawn from a Gaussian distribution with zero mean and variance,
\begin{eqnarray}\label{GD}
\overline{J_{i_1 i_2 i_3 i_4}^2} = \frac{6J^2}{N^{3}}\,,
\end{eqnarray}
where $J$ is a constant that sets the energy scale of the model.
\\

\noindent
In the large-$N$ limit, the low-energy dynamics of the SYK model are governed by the Schwarzian action, as established through the bulk-boundary correspondence \cite{Jensen:2016pah,Maldacena:2016upp,Engelsoy:2016xyb,Bagrets:2016cdf}.
This correspondence holds in the regime
\bea
G_2 \sim \frac{1}{N} \ll 1, \quad \phi_b \sim \frac{1}{\beta J} \ll 1\,,
\eea
where $\beta$ is the inverse temperature.
An additional matter-coupling term must be incorporated to capture the full coupling structure of JT gravity in this limit. Working in the large-$N$ framework makes it particularly convenient to analyze the bulk gravitational theory through its dual boundary description.

\subsubsection{Coupling of Scalar Field}
\noindent
We begin by considering the massless case, $m=0$, for the scalar field, which yields the conformal dimension
\bea
\Delta = \frac{1}{2} + \sqrt{\frac{1}{4} + \frac{m^2}{\Lambda}} = 1\,.
\eea
To reproduce the corresponding quadratic coupling terms of JT gravity, it is necessary to introduce a four-fermion interaction in the Majorana sector coupled to a scalar field. The resulting Hamiltonian takes the form \cite{Moitra:2022glw}
\bea
H_{\mathrm{SYK4}\phi} &=&
M\sum_{j=1}^M\frac{\pi_{\phi, j}^2}{2}\,\,
+\sum_{1\le i_1<i_2<i_3<i_4\le N} j_{i_1i_2i_3i_4}\psi_{i_1}\psi_{i_2}\psi_{i_3}\psi_{i_4}
\nn\\
&&
-\sum_{j=1}^M\sum_{1\le j_1<j_2<j_3<j_4\le N} g_{j_1j_2j_3j_4, j}\psi_{j_1}\psi_{j_2}\psi_{j_3}\psi_{j_4}\frac{\phi_j}{\sqrt{M}}\,.
\eea
The distribution of random couplings is given by
\bea
&&
\exp\Bigg(-\sum_{1\le i_1<i_2<i_3<i_4\le N}j_{i_1i_2i_3i_4}^2\frac{N^{3}}{12J^2}\Bigg)
\nn\\
&&\times
\exp\Bigg(-\sum_{j=1}^M\sum_{1\le j_1<j_2<j_3<j_4\le N}g_{j_1j_2j_3j_4, j}^2\frac{N^3}{2g^2}\Bigg),
\eea
where $g$ parameterizes the strength of the matter coupling.
Here, $\phi_j$ corresponds to the boundary value of the bulk scalar field $\Phi_j$, with the number of $\phi_j$ fields, $M$, matching the number of bulk scalar fields, $\widetilde{M}$.
\\

\noindent
Next, consider the lowest mass allowed in AdS spacetime, i.e., the Breitenlohner-Freedman bound \cite{Breitenlohner:1982bm},
\bea
\frac{m^2}{\Lambda} = \frac{1}{4}\,,
\eea
for which the conformal dimension is
\bea
\Delta = \frac{1}{2} + \sqrt{\frac{1}{4} - \frac{m^2}{\Lambda}} = \frac{1}{2}\,.
\eea
In this case, the interaction term is modified, yielding the Hamiltonian \cite{Moitra:2022glw}
\bea
H_{\mathrm{SYK2}\phi}
&=&
M\sum_{j=1}^M\frac{\pi_{\phi, j}^2}{2}
+\sum_{1\le i_1<i_2<i_3<i_4\le N} j_{i_1i_2i_3i_4}\psi_{i_1}\psi_{i_2}\psi_{i_3}\psi_{i_4}
\nn\\
&&\times\,\, i\sum_{j=1}^M\sum_{1\le i_1<i_2\le N} g_{i_1i_2, j}\psi_{i_1}\psi_{i_2}\frac{\phi_j}{\sqrt{M}}.
\eea
The corresponding random coupling distribution is
\bea
\exp\Bigg(-\sum_{1\le i_1<i_2<i_3<i_4\le N}j_{i_1i_2i_3i_4}^2\frac{N^{3}}{12J^2}\Bigg)
\times
\exp\Bigg(-\sum_{j=1}^M\sum_{1\le i_1<i_2\le N}g_{i_1i_2, j}^2\frac{N}{2g^2}\Bigg)\,.
\eea
By considering the following scaling relation between the coupling parameter and the SYK coupling,
\bea
\lambda \sim \frac{g^2}{J^3} \ll 1,
\eea
one can recover the coupling terms of JT gravity from the boundary theory in the large-$N$ limit.
A similar construction applies in the case of Dirac matter, where the conformal dimension guides the introduction of appropriate interaction terms between the Majorana fermions and the matter fields.

\subsubsection{Coupling of massless Dirac Matter}
\noindent
Next, we consider the case of massless Dirac matter, $m_{\mathrm{f}}=0$, for which the conformal dimension is,
\bea
\Delta = \frac{1}{2}\,.
\eea
The corresponding Hamiltonian is given by \cite{Lau:2023pot},
\bea
H_{\mathrm{SYK2}\chi}
&=&
\sum_{1\le i_1<i_2<i_3<i_4\le N} j_{i_1i_2i_3i_4} \psi_{i_1}\psi_{i_2}\psi_{i_3}\psi_{i_4}
\nn\\
&&
+\frac{i}{\sqrt{M}} \sum_{j=1}^{M} \sum_{1\le i_1<i_2\le N} \bigl( g_{i_1i_2, j} \psi_{i_1}\psi_{i_2} \chi_{j}^{\dagger}
+	g_{i_1i_2, j}^* \psi_{i_1}\psi_{i_2} \chi_{j} \bigr).
\label{HSYK}
\eea
The random couplings are drawn from the distribution \cite{Lau:2023pot}
\bea
\exp\Bigg(-\sum_{1\le i_1<i_2<i_3<i_4\le N} j_{i_1i_2i_3i_4}^2 \frac{N^{3}}{12J^2}\Bigg)
\times
\exp\Bigg(-\sum_{j=1}^M \sum_{1\le i_1<i_2\le N} g_{i_1i_2, j}^* g_{i_1i_2, j} \frac{N}{2g^2}\Bigg).
\nn\\
\eea
Since the bulk Dirac fermion is a two-component spinor $\Psi_j$, the number of boundary fields $\chi_j$ must satisfy $M = 2\widetilde{M}$ and therefore be even \cite{Lau:2023pot}. Notably, because the Hamiltonian contains an odd number of fermionic operators, it does not commute with the particle-hole operator \cite{Lau:2023pot}
\bea
\Gamma \equiv K \prod_{j=1}^{T_d}\bigl(c_j + c^{\dagger}_j\bigr),
\eea
where
\bea
T_d \equiv \frac{N}{2} + M,
\eea
$K$ is an anti-linear operator implementing complex conjugation, and the Dirac operators $c_j$ and $c^{\dagger}_j$ are considered real when $K$ acts on them. As a result, number-parity conservation is explicitly broken in this model \cite{Lau:2023pot}.

\subsubsection{Coupling of massive Dirac Matter}
\noindent
If, on the other hand, the Dirac fermion has mass, say $\mf$ with,
\bea
\frac{\mf}{\sqrt{-\Lambda}}=\frac{1}{4},
\eea
the associated conformal dimension is
\bea
\Delta=\frac{1}{2}+\frac{\mf}{\sqrt{-\Lambda}}=\frac{3}{4}.
\eea
The corresponding Hamiltonian then becomes
\bea
H_{\mathrm{SYK3}\chi} &=&
\sum_{1\le i_1<i_2<i_3<i_4\le N} j_{i_1i_2i_3i_4}\psi_{i_1}\psi_{i_2}\psi_{i_3}\psi_{i_4}
\nn\\
&&
+\frac{i}{\sqrt{M}}\sum_{j=1}^M\sum_{1\le i_1<i_2<i_3\le N} \big(g_{i_1i_2i_3, j}\psi_{i_1}\psi_{i_2}\psi_{i_3}\chi^{\dagger}_j
+g^*_{i_1i_2i_3, j}\psi_{i_1}\psi_{i_2}\psi_{i_3}\chi_j\big)
.\label{eqn:SYK3chi}
\nn\\
\eea
In this case, the distribution of the random couplings becomes
\bea
\exp\Bigg(-\sum_{1\le i_1<i_2<i_3<i_4\le N}j_{i_1i_2i_3i_4}^2\frac{N^{3}}{12J^2}\Bigg)
\times
\exp\Bigg(-\sum_{j=1}^M\sum_{1\le i_1<i_2\le N}g_{i_1i_2i_3, j}^*g_{i_1i_2i_3, j}\frac{N}{2g^2}\Bigg)\,.\label{eqn:SYK3chiCoupling}
\nn\\
\eea
\noindent
This Hamiltonian shares structural similarities with the doubled SYK model, which leads to the restoration of number-parity conservation.
It would be valuable to explore how the physical results depend on the value of the fermion mass since the mass term modifies the underlying symmetry structure.
The gravitational theory emerges in the regime where the matter coupling parameter satisfies the scaling relation
\bea
\lambda \sim \frac{g^2}{J^2} \ll 1,,
\eea
reflecting the correspondence between the gravity parameter $\lambda$ and the SYK model parameters.

\medskip

\noindent
Up to this point, we have focused exclusively on quadratic matter couplings.
Quartic and higher-order matter couplings do not persist in the large-$N$ limit under standard assumptions.
However, introducing a non-Gaussian disorder distribution will show how quartic matter couplings can survive.

\subsubsection{Quartic Matter Coupling}
\noindent
To demonstrate the scalar field case with a minimally allowed mass value, we modify the distribution by introducing a non-Gaussian distribution,
\bea
&&
\exp\Bigg(-\sum_{1\le i_1<i_2<i_3<i_4\le N}j_{i_1i_2i_3i_4}^2\frac{N^{3}}{12J^2}\Bigg)
\times
\exp\Bigg(-\sum_{j=1}^M\sum_{1\le i_1<i_2\le N}g_{i_1i_2, j}^2\frac{N}{2g^2}\Bigg)
\nn\\
&&\times
\exp\Bigg(-\sum_{j=1}^M\sum_{1\le i_1,i_2<k_1, k_2\le N}g_{i_1k_1, j}g_{i_1k_2, j}g_{i_2k_1, j}g_{i_2k_2, j}\frac{3N}{4g^4}\Bigg)\,,
\label{eqn:quarticDistribution}
\eea
leaving Hamiltonian unchanged.
When the disorder parameters follow a Gaussian distribution, the random coupling constants can be integrated exactly. However, in the presence of non-Gaussian disorder parameters, one must instead perform a small-$g$ expansion to carry out the integration consistently.
\\

\noindent
To leading order in $1/N$, this procedure is equivalent to varying the couplings $g_{i_1, i_2, j}$ and extracting perturbative solutions:
\bea
g_{i_1i_2, j}=g^{(1)}_{i_1i_2,j}+g^{(2)}_{i_1i_2, j}+\cdots.
\eea
The explicit solution up to the second order is given by:
\bea
g^{(1)}_{i_1i_2, j}=i\frac{g^2}{\sqrt{M}}\int d\tau\ \frac{\psi_{i_1}(\tau)\psi_{i_2}(\tau)}{N}\phi_j(\tau);
\eea
\bea
g^{(2)}_{i_1i_2, j}=-\frac{3}{g^2}\sum_{1\le i_1, j_2<i_2, i_3\le N}g^{(1)}_{i_1i_3, j}g^{(1)}_{j_2 i_2, j}g^{(1)}_{j_2 i_3, j}
.
\eea
Substituting this perturbative solution into the random distribution and the Hamiltonian, we obtain the low-energy effective action,
\bea
\frac{S_{\mathrm{eff}}}{N}
&=&-\frac{g^2}{2!}\frac{1}{2}\int d\tau_1d\tau_2\ \frac{\sum_{i_1=1}^N\psi_{i_1}(\tau_1)\psi_{i_1}(\tau_2)}{N}\frac{\sum_{i_2=1}^N\psi_{i_2}(\tau_1)\psi_{i_2}(\tau_2)}{N}\frac{\sum_{j=1}^M\phi_j(\tau_1)\phi_j(\tau_2)}{M}
\nn\\
&&
-3\frac{g^4}{4!}\int d\tau_1d\tau_2d\tau_3d\tau_4\ \frac{\sum_{i_1=1}^N\psi_{i_1}(\tau_1)\psi_{i_1}(\tau_2)}{N}\frac{\sum_{i_2=1}^N\psi_{i_2}(\tau_1)\psi_{i_2}(\tau_3)}{N}
\nn\\
&&\times
\frac{\sum_{i_3=1}^N\psi_{i_3}(\tau_2)\psi_{i_3}(\tau_4)}{N}\frac{\sum_{i_4=1}^N\psi_{i_4}(\tau_3)\psi_{i_4}(\tau_4)}{N}
\frac{\sum_{j=1}^M\phi_j(\tau_1)\phi_j(\tau_2)\phi_j(\tau_3)\phi_j(\tau_4)}{M^2}
\nn\\
&&+\cdots.
\eea
The first and second terms correspond to the one-loop contributions from the connected two-point and four-point functions of $\phi_j$, respectively.
Considering a fully non-Gaussian disorder distribution amounts to taking the large-$N$ limit while resumming the one-loop contributions of all connected correlators — effectively summing over all orders in $g$.
In this work, we focus on the quartic disorder distribution.
Extending the analysis to include terms of order $g^6$, corresponding to the connected six-point function, would require introducing a sextic distribution.
For Dirac matter, a similar treatment of the disorder distribution can be applied without modifying the Hamiltonian, as in the bosonic case.
\\

\begin{figure}
\centering
\includegraphics[height=12cm]{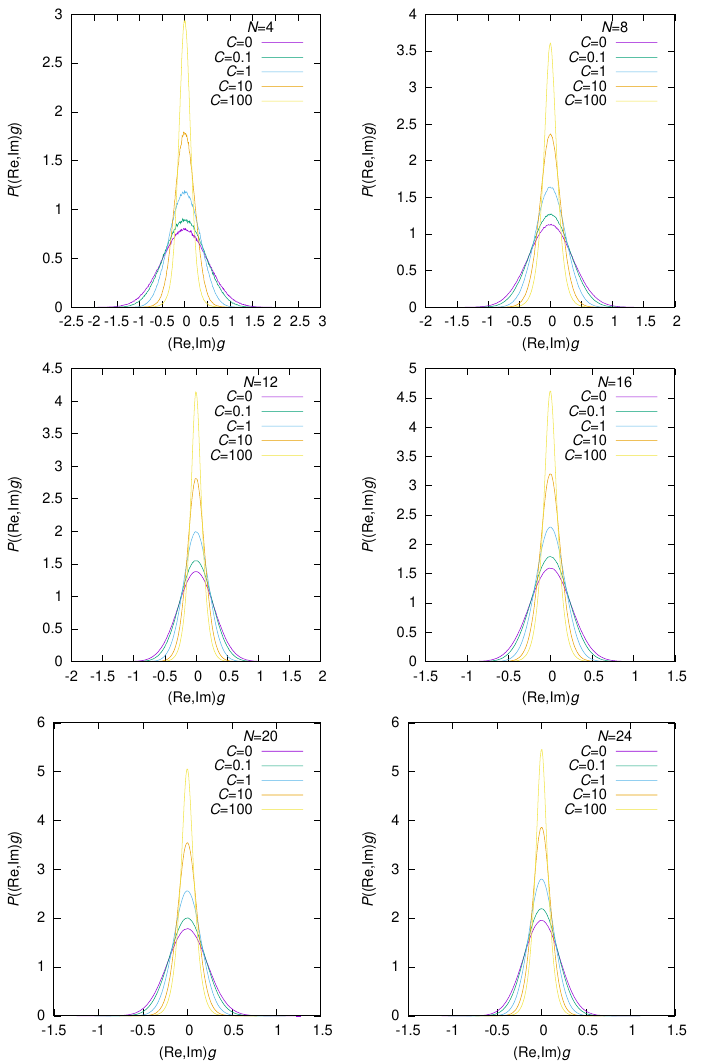}
\caption{Probability distribution for the coupling $g_{i_1i_2,j}$ \eqref{eqn:quarticDistribution2} for various $N$ and $C$. The distribution does not depend on the choice of $M$. $10^5$ samples are collected with $10^4$ Monte Carlo steps between each sample.
The horizontal axis is the real part or imaginary part of the coupling, which shows the same distribution.
}
\label{fig:correlatedCouplings}
\end{figure}

\noindent
In analogy to Eq. \eqref{eqn:quarticDistribution}, we consider a distribution
\bea
&&
\exp\Bigg(-\sum_{1\le i_1<i_2<i_3<i_4\le N}j_{i_1i_2i_3i_4}^2\frac{N^{3}}{12J^2}\Bigg)
\nn\\
&&\times
\exp\Bigg(-\sum_{j=1}^M\sum_{1\le i_1<i_2\le N}|g_{i_1i_2, j}|^2\frac{N}{2g^2}\Bigg)
\nn\\
&&\times
\exp\Bigg(-C\sum_{j=1}^M\sum_{1\le i_1,i_2<k_1, k_2\le N} g^*_{i_1k_1, j}g^*_{i_1k_2, j}g_{i_2k_1, j}g_{i_2k_2, j} \frac{3N}{4g^4}\Bigg)\,,
\label{eqn:quarticDistribution2}
\eea
for the quartic Dirac matter coupling, $g_{i_1 k_2, j}$. Here, $C$ is a positive constant that controls the strength of the non-Gaussianity. As $C$ increases, the distribution of $g_{i_1i_2,j}$ progressively deviates from the Gaussian form. Figure~\ref{fig:correlatedCouplings} demonstrates that this non-Gaussian behavior is not suppressed in the large-$N$ limit.


\section{Quantum Chaos and Entanglement}
\label{sec:3}
\noindent
In this section, we will focus on the coupling effects in the Dirac matter case. Unlike the scalar field case, it allows for numerically tractable analysis without requiring artificial regularization.
To probe signatures of quantum chaos, we evaluate the adjacent gap ratio \cite{Atas:2012prl} and the spectral form factor (SFF) \cite{Brezin:1997rze}, which together characterize short- and long-range spectral correlations.
In addition, we study quantum entanglement by computing the entanglement entropy between the Majorana and Dirac fermions, allowing us to extract the associated von Neumann algebra structure.
Finally, we explore the broader holographic implications for quantum chaos and entanglement dynamics by incorporating both non-Gaussian disorder distributions and bulk fermion mass effects.

\subsection{Quantum Chaos}
\noindent
In the following, we demonstrate that the Hamiltonian's symmetry alone is sufficient to classify the associated random matrix ensemble, even when the random coupling distributions are modified.
We then analyze the spectral form factor (SFF).
In a quantum chaotic model, the SFF typically exhibits a characteristic dip-ramp-plateau structure \cite{Cotler:2016fpe}.
Initially, destructive interference between energy levels causes the SFF to decrease (the dip).
As time evolves, the SFF displays a linear ramp, reflecting correlations among energy levels, before finally reaching a plateau, indicating the saturation of correlations.
Our study shows that non-Gaussian disorder extends the timescale of the linear ramp, delaying the onset of the plateau. At the same time, the inclusion of a bulk fermion mass shortens the plateau timescale. These shifts in the long-range spectral correlations provide a way to probe both non-Gaussian disorder effects and bulk fermion mass contributions.

\subsubsection{Adjacent Gap Ratio}

\noindent
To analyze short-range spectral correlations, we compute the adjacent gap ratio
\bea
\{r_j\}_{j=1}^{D-2}=\left\{\frac{\min(\delta_j,\delta_{j+1})}{\max(\delta_j,\delta_{j+1})}\right\}_{j=1}^{D-2},
\label{eqn:r}
\eea
where the eigenvalue gaps are defined by identifying the non-degenerate eigenvalues $$\{E_1, E_2, \ldots, E_D\}$$ in each parity sector and computing
\bea
\delta_1 = E_2 - E_1,\quad \delta_2 = E_3 - E_2, \quad \ldots, \quad \delta_{D-1} = E_D - E_{D-1}\,.
\eea
This ratio is a well-established diagnostic that distinguishes between random matrix ensembles, such as the Gaussian orthogonal ensemble (GOE), the Gaussian unitary ensemble (GUE), and the Gaussian symplectic ensemble (GSE), characterized by their symmetries. The known ensemble averages are approximately 0.5307 for GOE, 0.59975 for GUE, and 0.6744 for GSE \cite{Atas:2012prl,Nishigaki:2024yjr}.
The charge-conjugation operators
\bea
{\cal P} \equiv K \prod_{j=1}^{T_d} \gamma_{2j-1}, \quad
{\cal R} \equiv K \prod_{j=1}^{T_d} i \gamma_{2j},
\label{cc}
\eea
written in terms of the Majorana operators $\gamma_j$, continue to commute with the Hamiltonian, preserving its symmetry classification under random matrix theory.

\medskip

\noindent
We particularly focus on the Hamiltonian \eqref{HSYK}, which violates the conservation of number parity.
Its first term involves a four-Majorana product, while the second includes a two-Majorana product.
The number parity of complex fermions formed from these Majorana modes remains conserved, leading to a block-diagonal structure in the matrix representation of Eq. \eqref{HSYK}. Effectively, the parity operator behaves similarly to the standard SYK model but acts trivially on the matter sector \cite{Lau:2023pot}.
Therefore, we define modified parity operators for Eq. \eqref{HSYK} to classify the system’s symmetries correctly.

\medskip

\begin{figure}
\includegraphics[width=15cm]{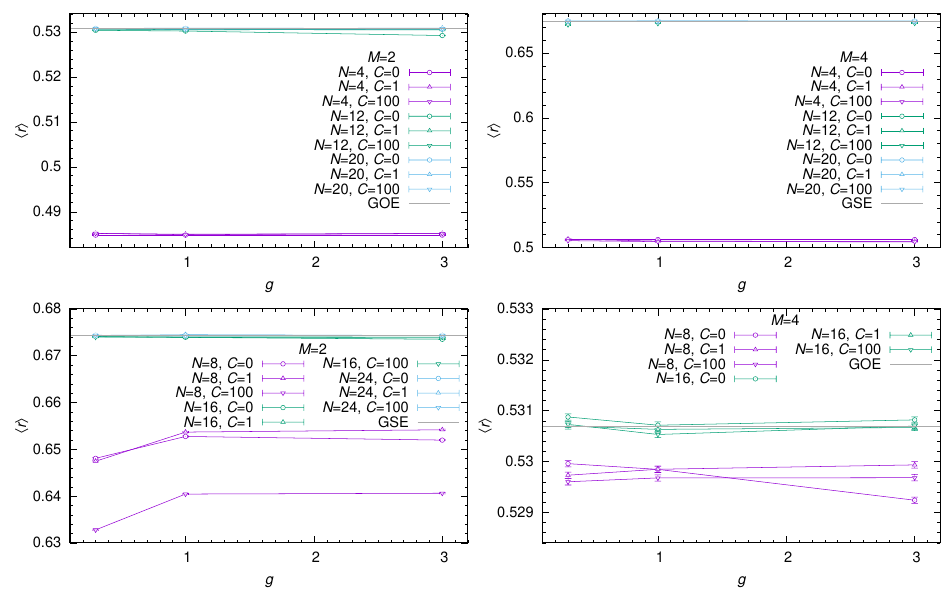}
\caption{Averaged adjacent gap ratio for the entire energy spectrum after removing degeneracy.
$2^{24}$ eigenvalues are used.
Here and thereafter, $10^{4}$ MC steps are taken between each sample for $C>0$.}
\label{fig:adjacent_gap_ratio}
\end{figure}

\noindent
For the non-Gaussian distribution above, the form of the Hamiltonian remains unchanged, so the spectral degeneracies are preserved, consistent with previous observations \cite{Lau:2023pot}.
As shown in Fig.~\ref{fig:adjacent_gap_ratio}, however, the average value of the adjacent gap ratio changes, even though the symmetry classification itself does not.
This observation highlights a subtle but nontrivial effect; for sufficiently large $N/M$, the degeneracy patterns and random matrix classes can be summarized as follows:
\begin{itemize}
\item $T_d \bmod 4 = 0, ({\cal P}^2, {\cal R}^2) = (1, 1):$ no degeneracy, GOE;
\item $T_d \bmod 4 = 1, ({\cal P}^2, {\cal R}^2) = (1, -1):$ no degeneracy, GUE;
\item $T_d \bmod 4 = 2, ({\cal P}^2, {\cal R}^2) = (-1, -1):$ twofold degeneracy, GSE;
\item $T_d \bmod 4 = 3, ({\cal P}^2, {\cal R}^2) = (-1, 1):$ no degeneracy, GUE.
\end{itemize}
Since the adjacent gap ratio is sensitive only to short-range energy correlations, the symmetry-based classification holds even in the presence of modified disorder statistics.
In the next section, we will see how long-range correlations captured by the SFF reveal differences beyond these short-range properties.

\subsubsection{Spectral Form Factor}
\noindent
Another increasingly popular diagnostic of quantum chaos based on the energy spectrum is the spectral form factor (SFF), which is defined as the square of the Fourier transform of the empirical spectral density \cite{Brezin:1997rze,Dyer:2016pou}. After performing the disorder average, we define the total SFF ($g$), as well as its disconnected part ($g_d$) and connected part ($g_c$), as follows:
\begin{align}
g(t,\beta) &\equiv \frac{\langle Z(\beta,t), Z^{*}(\beta,t)\rangle}{\langle Z(\beta)\rangle^{2}}\,, \\
g_{d}(t, \beta) &\equiv \frac{\langle Z(\beta,t)\rangle\cdot \langle Z^{*}(\beta,t)\rangle}{\langle Z(\beta)\rangle^{2}}\,, \\
g_{c}(t, \beta) &\equiv g(t, \beta) - g_{d}(t, \beta)\,,
\end{align}
where the partition function
\bea
Z(\beta,t) \equiv \mathrm{Tr}\left(e^{-\beta H - i H t}\right)\,,
\eea
traces over the time-evolved and thermally weighted Hamiltonian.
The notation $\langle \cdot \rangle$ indicates averaging over an ensemble of random couplings.
\\



\begin{figure}
\includegraphics{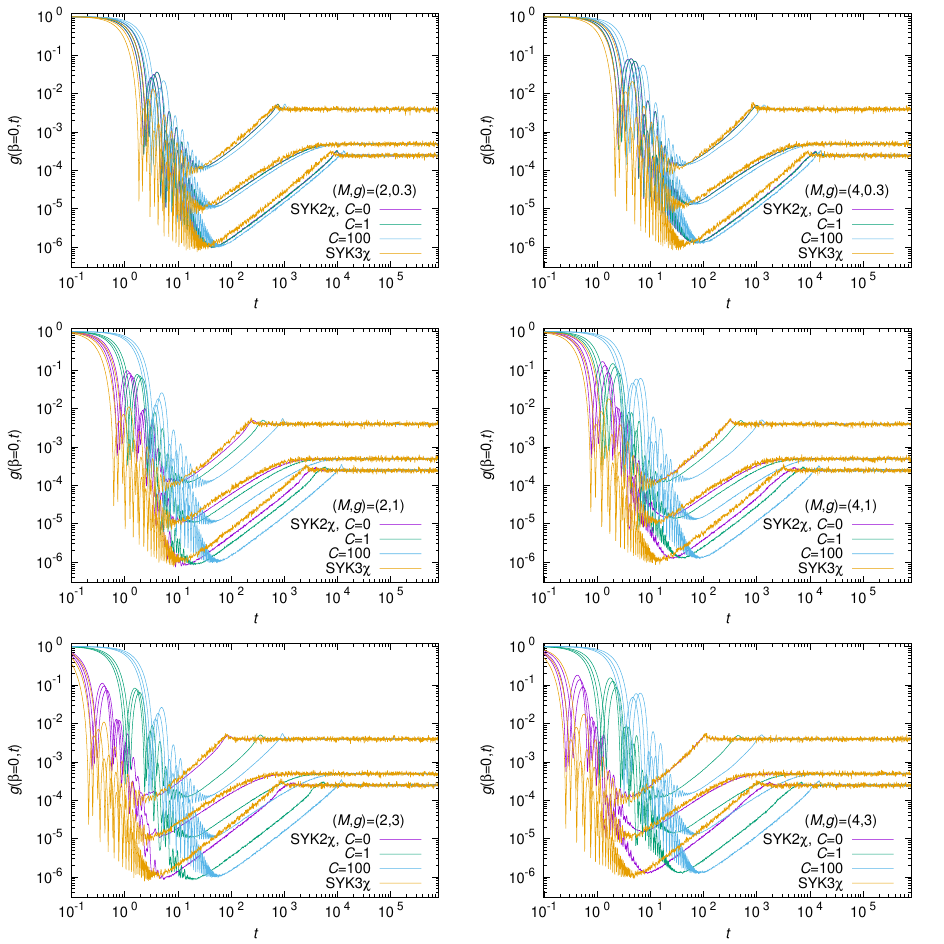}
\caption{
Spectral form factor $g(t)$ for SYK$2\chi$ \eqref{HSYK} and SYK$3\chi$ \eqref{eqn:SYK3chi}.
Only the contribution of the even-parity sector has been computed for simplicity.
For $M=2$, results for $N=16,20,24$ are plotted.
For $M=4$, results for $N=12,16,20$ are plotted.
The late-time values decrease as $N$ is increased.
$2^{25-N/2-M}$ samples are used for SYK$2\chi$ while 128 samples are used for SYK$3\chi$.
}
\label{fig:SFFM3D1vsM2D1}
\end{figure}

\begin{figure}
\includegraphics{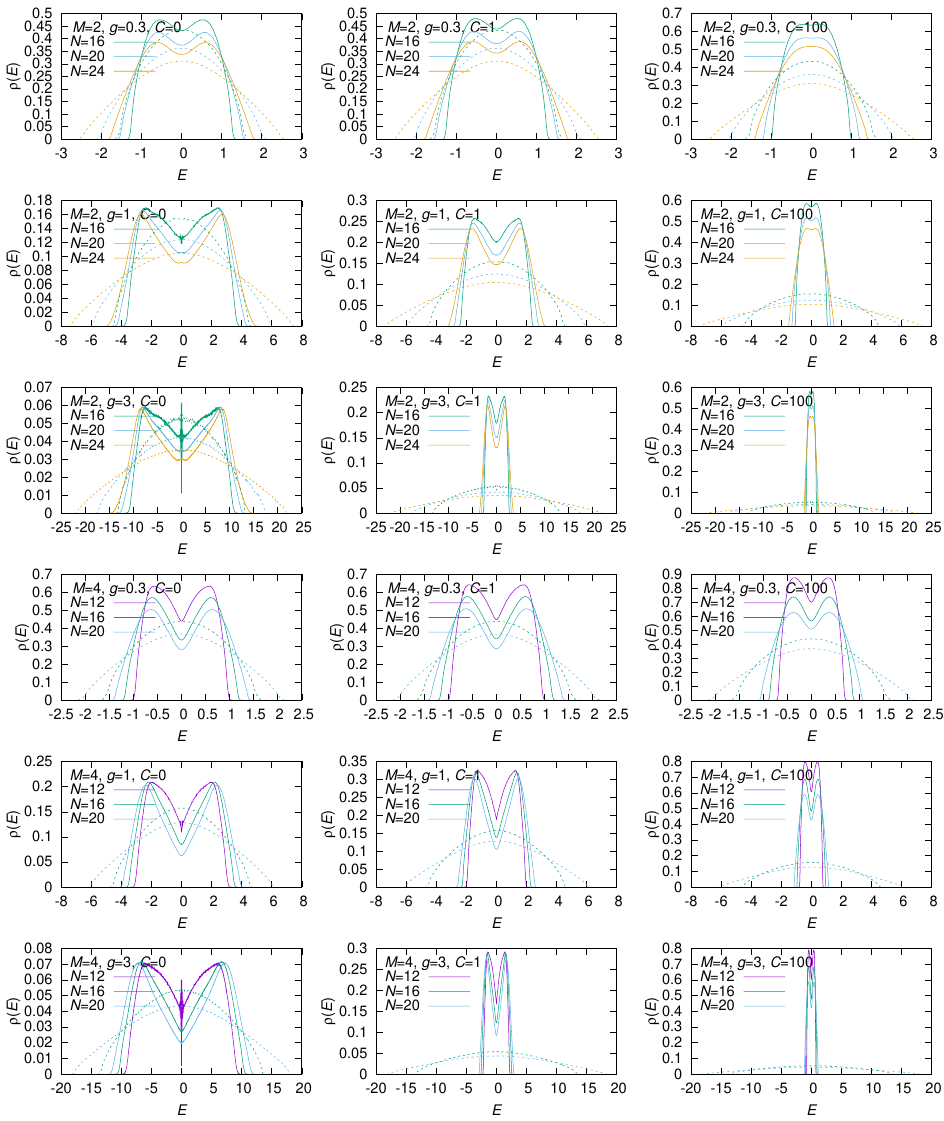}
\caption{
Normalized density of states ($\int \rho(E)dE=1$) for SYK2$\chi$ \eqref{HSYK} (solid line) and SYK$3\chi$ \eqref{eqn:SYK3chi} (dashed line).
$2^{25-N/2-M}$ ($2^{20-N/2-M}$) samples are used for SYK$2\chi$ (SYK$3\chi$).
}
\label{fig:dos}
\end{figure}

\noindent
In Fig.~\ref{fig:SFFM3D1vsM2D1}, we compare the spectral form factor of Eq. \eqref{eqn:SYK3chi} against that of Eq. \eqref{HSYK}.
Here, the spectral form factor is computed using the even-parity sector of the Hamiltonian $\mathcal{H}_\mathrm{e}$:
\bea
g(t,\beta)=\frac{\langle Z(\beta,t)Z^{*}(\beta,t)\rangle}{\langle Z(\beta)\rangle^{2}};
Z(\beta,t)\equiv\mathrm{Tr}\left(e^{-(\beta+it)\mathcal{H}_\mathrm{e}}\right).
\eea
The timescale associated with the linear ramp is modified, as confirmed by the density of states (Fig.~\ref{fig:dos}). Because a non-Gaussian disorder distribution extends the equilibration timescale, we anticipate that incorporating {\it all} non-Gaussian contributions in the large-$N$ limit — particularly when the coupling strength exceeds a critical threshold — could push the equilibrium time to infinity.
In this regime, the breakdown of the equilibrium state suggests that the matter sector effectively suppresses the gravitational dynamics.

\subsection{Entanglement Entropy and von Neumann Algebra}
\noindent
Numerical investigations of entanglement entropy usually start by partitioning a given system into two subsystems, denoted $A$ and $B$, and computing
\bea
S_{\mathrm{EE}} = -\mathrm{Tr}_A \left( \rho_A \ln \rho_A \right),
\eea
where $\mathrm{Tr}_A$ denotes the partial trace over subsystem $A$, and $\rho_A$ is the reduced density matrix on $A$.
In our case, the total Hilbert space factorizes as $\mathscr{H}_M \otimes \mathscr{H}_D$, where $\mathscr{H}_M$ is associated with the Majorana SYK model and $\mathscr{H}_D$ with the Dirac matter sector.
We perform the partial trace over the matter sector to evaluate the entanglement entropy, starting from the unoccupied initial state $\ket{0 \cdots 0}$ in the occupation number basis.

\medskip

\begin{figure}
\centering\includegraphics{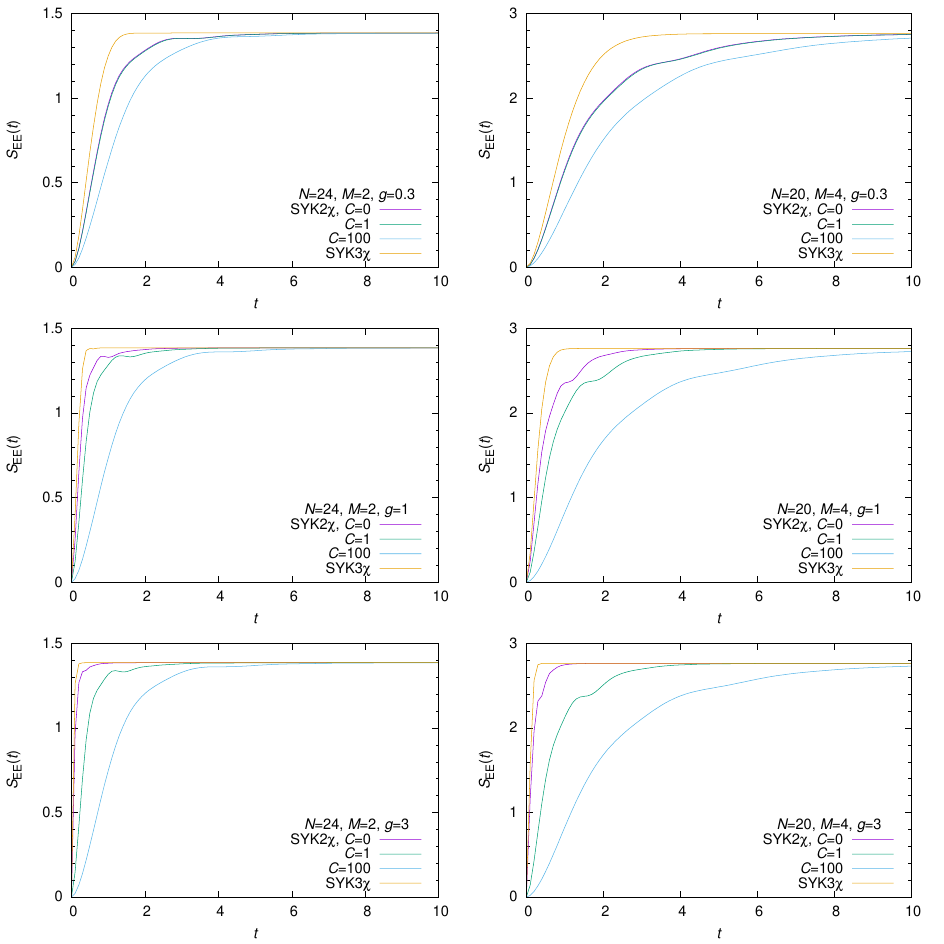}
\caption{
Dynamical behavior of the averaged entanglement entropy for SYK$2\chi$ and SYK$3\chi$.
$2^{20-N/2-M}$ samples are used.
}
\label{fig:EE-early}
\end{figure}

\begin{figure}
\includegraphics{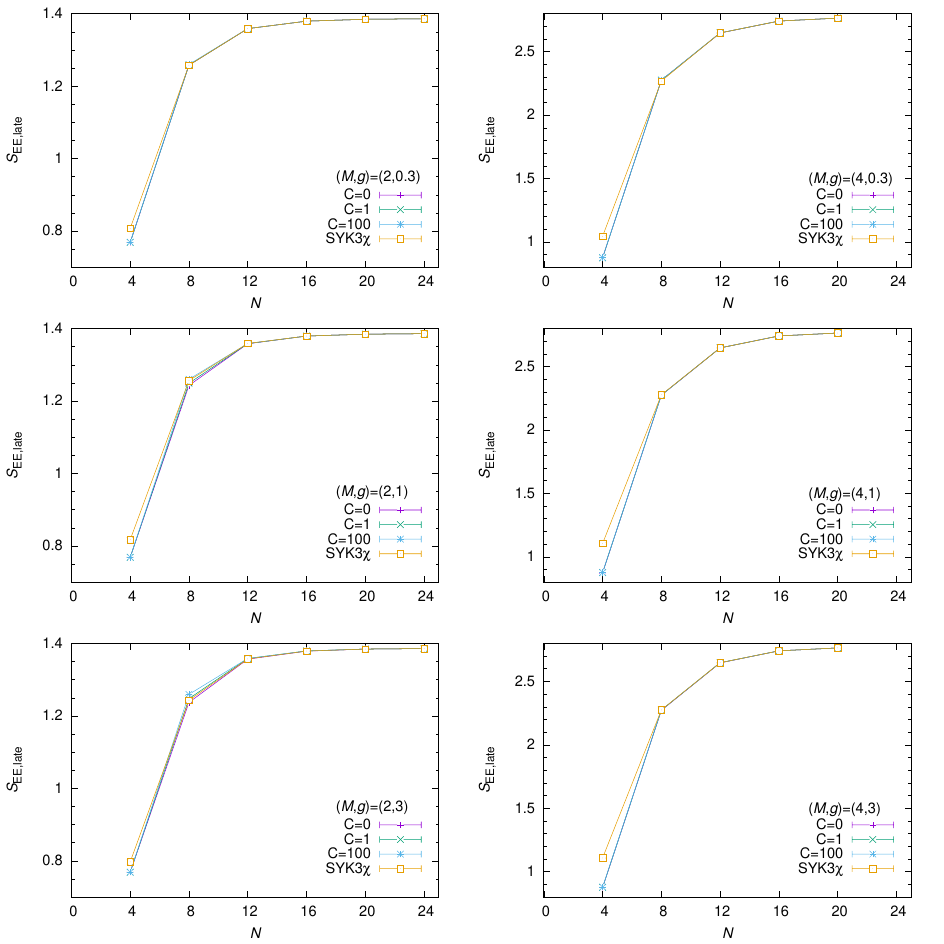}
\caption{
Late-time behavior of the averaged entanglement entropy for SYK$2\chi$ and SYK$3\chi$.
The average of results for $t=(1,2,\ldots,10)\times10^6$ is plotted.
$2^{20-N/2-M}$ samples are used.
}
\label{fig:EE-late}
\end{figure}

\noindent
For the Hamiltonian defined in Eq. \eqref{eqn:SYK3chi} with random couplings sampled according to Eq. \eqref{eqn:SYK3chiCoupling}, we numerically examine the time evolution of $S_{\mathrm{EE}}$ between the $N$ Majorana fermions and the $M$ Dirac fermions.
Specifically, we analyze 128 disorder realizations and compute $S_{\mathrm{EE}}(t)$ for time points
$t = 0,\, \Delta t,\, 2 \Delta t,\, \ldots,\, N_t \Delta t$
with $\Delta t = 0.1$ and $N_t = 100$, unless otherwise stated. The results are presented in Fig. \ref{fig:EE-early} for the early-time dynamics, and in Fig. \ref{fig:EE-late} for a comparison of late-time entanglement entropy across different models: the SYK${2\chi}$ model \eqref{HSYK} with either correlated ($C=100$) or uncorrelated ($C=0$) couplings \eqref{eqn:quarticDistribution2}, and the SYK${3\chi}$ model \eqref{eqn:SYK3chi} with uncorrelated couplings \eqref{eqn:SYK3chiCoupling}.

\medskip

\noindent
Our numerical results suggest:
\begin{itemize}
\item The entanglement entropy grows initially and then saturates;
\item In the large-$N$ limit, it saturates to the maximal possible value $M \ln 2$;
\item The saturation time remains finite in the large-$N$ limit. However, it is sensitive to non-Gaussian disorder and the bulk mass effects.
\end{itemize}
These findings indicate that the state is well-approximated by a maximally entangled state between the Majorana fermions and Dirac matter fields.
Invoking the Schmidt decomposition theorem, the state after reaching the saturation regime in the large-$N$ limit can be expressed as
\bea
\ket{\psi} = \frac{1}{\sqrt{2^M}} \sum_{j=1}^{2^M} \ket{j_M} \otimes \ket{j_D}\,,
\eea
where $\{\ket{j_M}\}$ and $\{\ket{j_D}\}$ form orthonormal bases for $\mathscr{H}_M$ and $\mathscr{H}_D$, respectively.
\\

\noindent
In the saturation regime, the algebra of observables transitions from a von Neumann type I algebra to a type II algebra as $N \rightarrow \infty$.
Our findings indicate that, according to the low-energy effective theory, the graviton becomes maximally entangled with the Dirac fermion matter fields once $N$ is sufficiently large. The emergence of a maximally mixed state implies that the Majorana fermion algebra realizes a type II$_1$ structure in the limit $N = 2M \rightarrow \infty$.
Since the subalgebra is independent of the choice of partial trace region, the global algebra of the large-$N$ theory is equivalent to a von Neumann type II$_1$ algebra in the saturation regime.
This algebraic transition from type I to type II reflects a form of information loss.

\medskip

\noindent
Numerically, we observe the maximally mixed state in the saturation regime at large $N$ via the entanglement entropy. Notably, non-Gaussian disorder requires a longer timescale to reach this equilibrium state.
This is particularly interesting, as introducing all bulk matter couplings could destabilize the equilibrium when the matter coupling strength exceeds a critical threshold.
In such a scenario, the disappearance of the maximally mixed state would signal the emergence of a type II$_{\infty}$ algebra on finite timescales.
Therefore, truncating the bulk matter couplings to achieve a type II$_1$ structure or a rapid saturation time should be interpreted as an artifact of the modeling assumptions.

\medskip

\noindent
Furthermore, introducing a bulk fermion mass reduces the time required to reach equilibrium.
In principle, one might expect a similar transition behavior when the fermion mass is varied continuously.
However, our current model does not support a continuum of fermion mass values, making it difficult to draw firm conclusions about the dynamic transitions in that parameter space.
\newpage

\section{Discussion and Conclusion}
\label{sec:4}
\noindent
This paper investigated the coupling of Dirac matter fields in the SYK model, focusing on its implications for holographic and quantum information.
By introducing interaction terms coupling Majorana fermions to Dirac matter fields, we extended the usual SYK framework \cite{Polchinski:2016xgd} to probe how matter fields influence the emergent gravitational description in JT gravity.
In particular, we generalized the standard Gaussian disorder average to incorporate bulk matter couplings, which required handling non-Gaussian distributions within the large-$N$ limit. While this introduces non-local effects \cite{Gross:2017hcz}, we argued that large-$N$ techniques can still yield a tractable, effectively local bulk description — crucial for connecting to field-theoretic approaches in holography.
\\

\noindent
Our analysis of quantum chaos properties of the model employed both the adjacent gap ratio \cite{Atas:2012prl}, which diagnoses short-range spectral correlations, and the spectral form factor (SFF) \cite{Brezin:1997rze}, which encodes long-range correlations.
We found that non-Gaussian disorder distributions and bulk matter couplings do not change the SYK model's symmetry classification, maintaining its consistency with random matrix theory.
However, they influence dynamical properties such as the timescale for reaching equilibrium, with stronger bulk couplings potentially delaying or eliminating the SFF plateau.
These observations highlight how long-range correlations are sensitive probes of bulk matter effects while short-range classifications remain robust.
\\

\noindent
We further examined entanglement entropy as a probe of emergent spacetime.
In the large-$N$ regime, the coupling of graviton-like modes to Dirac matter produced maximally mixed states corresponding to a type II$_1$ von Neumann algebra \cite{Lau:2023pot}.
Including more bulk matter couplings lengthens equilibration times and eventually destroys the maximally mixed state, indicating a transition toward a type II$_\infty$ algebra on finite timescales.
This suggests that the vN II$_1$ structure commonly found in the large-$N$ limit is an artifact of neglecting higher-order matter interactions. The more correct algebraic picture involves vN II$_\infty$.
Such results are valuable for understanding the role of non-Gaussian disorder and the deeper algebraic structure of holographic duals.
\\

\noindent
While we focused on quartic-order bulk matter couplings, the methodology can be extended to higher orders.
Future work could confirm whether these observations persist beyond the current approximation.
Although we focused on Dirac matter due to its analytic advantages, extending the framework to bosonic matter would be interesting, despite the additional regularization challenges, as it could reveal new features of holographic quantum chaos and entanglement.
\\

\noindent
We also briefly noted broader implications for holography and cosmology.
In particular, the emergence of type II$_1$ von Neumann factors resonates with proposals regarding de Sitter entropy and the structure of cosmological horizons, suggesting that techniques from quantum information theory could shed light on spacetime dynamics.
These connections point to a consistent picture in which higher-dimensional bulk geometries and their thermodynamic properties emerge from lower-dimensional boundary theories, consistent with the holographic principle.
Although lower-dimensional, the SYK model serves as a functional toy laboratory for exploring these ideas before applying them to more realistic, higher-dimensional gravity theories.
\\

\noindent
Finally, we offered a perspective on integrability in disordered systems.
Despite exhibiting chaotic behavior, the SYK model also displays features reminiscent of integrability in its random coupling structure.
This challenges the traditional dichotomy between integrable and non-integrable systems, motivating the refinement of the definition of integrability in ensemble-averaged settings \cite{Muller:2004nb}.
The adjacent gap ratio remains a powerful tool in this regard, and future investigations of different random coupling statistics could deepen our understanding of how integrability and chaos coexist in holographic models.

\section*{Acknowledgments}
\noindent 
We want to express our gratitude to Yifan Chen, Xing Huang, and Yikun Jiang for their helpful discussion. 
PHCL would like to acknowledge support from JSPS KAKENHI Grant No.~JP23H01174.
JM would like to acknowledge support from the ``Quantum Technologies for Sustainable Development''  grant from the National Institute for Theoretical and Computational Sciences of South Africa (NITHECS).
MT acknowledges the Grants-in-Aid from MEXT of Japan (Grants No. JP20K03787, JP21H05185, and JP25K00925) and JST CREST (Grant No. JPMJCR24I2). 
CTM thanks Nan-Peng Ma for his encouragement. 
The authors would like to thank the Isaac Newton Institute for Mathematical Sciences, Cambridge, for support and hospitality during the programme Quantum field theory with boundaries, impurities, and defects, where work on this paper was undertaken. 
This work was supported by EPSRC grant EP/Z000580/1.


\end{document}